\documentclass[amsmath,twocolumn,showpacs,aps,prb,superscriptaddress]{revtex4}
\usepackage{bm}
\usepackage{graphicx}
\usepackage{subfigure}
\newcommand{\bq}{\begin{equation}}
\newcommand{\ee}{\end{equation}}
\begin{document}

\title{Suppression of diffusion of hydrogen adatoms on graphene by effective adatom interaction}

\author{J. Talbot}
\affiliation{Department of Chemistry, University of Utah, Salt Lake City, UT 84112, USA}

\author{S. LeBohec}
\affiliation{Department of Physics and Astronomy, University of Utah, Salt Lake City, UT 84112, USA}

\author{E. G. Mishchenko}
\affiliation{Department of Physics and Astronomy, University of Utah, Salt Lake City, UT 84112, USA}

\begin{abstract}
Resonant graphene dopants, such as hydrogen adatoms, experience long-range effective interaction mediated by conduction electrons. As a result of this interaction,  when several adatoms are present in the sample, hopping of adatoms between sites belonging to different sublattices involves significant energy changes. Different inelastic mechanisms facilitating such hopping -- coupling to phonons and conduction electrons -- are considered. It is estimated that the diffusion of hydrogen adatoms is rather slow, amounting to roughly one hop to a nearest neighbor per millisecond.
\end{abstract}

 \maketitle

\section{Introduction}

 Graphene is a two-dimensional
material\cite{CN} holding a lot of technological potential.
Diffusion of hydrogen on graphene is a problem of major
interest for a number of applications. One potential application  arises from the
possibility of using graphene for hydrogen storage. For its successful implementation
it is important to understand and be able to predict the rates
of hydrogen absorption, desorption, and diffusion on graphene.

A different class of applications concerns graphene electronics. One
of the main obstacles here is graphene's good electric conduction in the
intrinsic state. It is thus desirable to develop ways of suppressing
graphene conductivity and  turning graphene into a semiconductor in a
controllable way. Avenues explored to achieve this objective
 include a number of possibilities: opening a
gap in graphene bilayers with an interlayer bias
\cite{CF,KCM,ZTG,MLS}, applying elastic strain
\cite{NWM,NYL,SN,KZJ,PCP,CCC}, carving out finite-width
nanoribbons \cite{E,SCL,HOZ}, inducing strong spin-orbital
coupling \cite{KM,QYF,WHA}, or using chemical doping
\cite{BME,ENM,BJN,DQF}.

Hydrogen is one especially promising dopant. Complete coverage
of graphene  with hydrogen atoms, however, results in
a dielectric (graphane) with a very large gap, $\sim 5\,
\text{eV}$, see Refs.~\onlinecite{SCB}, \onlinecite{ENM}, a situation equally
unfavorable for electronics applications. Nonetheless, partially
hydrogenated graphene remains a viable candidate. But for
an incomplete coverage of graphene with hydrogen, the
question of  diffusion becomes important. This question is
made much more interesting and non-trivial by the existence of an effective
interaction between the hydrogen atoms. Let us briefly explain the
origin of this interaction before discussing how it might
affect hydrogen transport.

A hydrogen impurity is resonant; its spectrum has an energy
level close to the Dirac point of the  conduction $\pi$-band of
graphene \cite{WKL}. Resonant hopping of
conduction electrons on and off the hydrogen
level gives rise to a large scattering amplitude of conduction electrons. It turns out that resonant scattering leads to an
 effect quite similar to that of a vacancy (a lattice site rendered unaccessible to
$\pi$-electrons because of the lack of a carbon atom there), or to a very strong potential of a substitution defect. As a result, the wave functions of
electrons are significantly modified. This  leads to a long-range interaction between dopants mediated by conduction electrons. In its origin, this interaction  is similar to the RKKY interaction or the classic Casimir
effect mediated by  virtual photons \cite{MT}. One notable difference with the RKKY interaction is that the latter is
usually considered for distances exceeding the Fermi
wavelength, $k_FR \gg 1$, while in the case of intrinsic graphene
one has to deal with the opposite limit, $k_FR \ll 1$.

For vacancies and also for strong substitution defects or resonant
impurities, the interaction energy $W(R)$ of two
impurities can already be estimated from dimensional considerations.
Graphene $\pi$-electrons have a  gapless Dirac spectrum,
$E=\pm vp$. Thus, for infinitely strong impurities in an intrinsic
graphene, one can construct only a single combination with the dimension of
energy: $W(R) \sim \hbar v/R$. This interaction
was first found in Ref.~\onlinecite{SAL} for $k_F=0$. Its dependence
on the chemical potential $\mu=v p_F$ was elucidated in
Ref.~\onlinecite{LTM}. Due to the presence of two sublattices in a
honeycomb graphene lattice and the resulting quantum
interference,  the sign of the interaction depends on whether
\begin{figure}[h]
\resizebox{.37\textwidth}{!}{\includegraphics{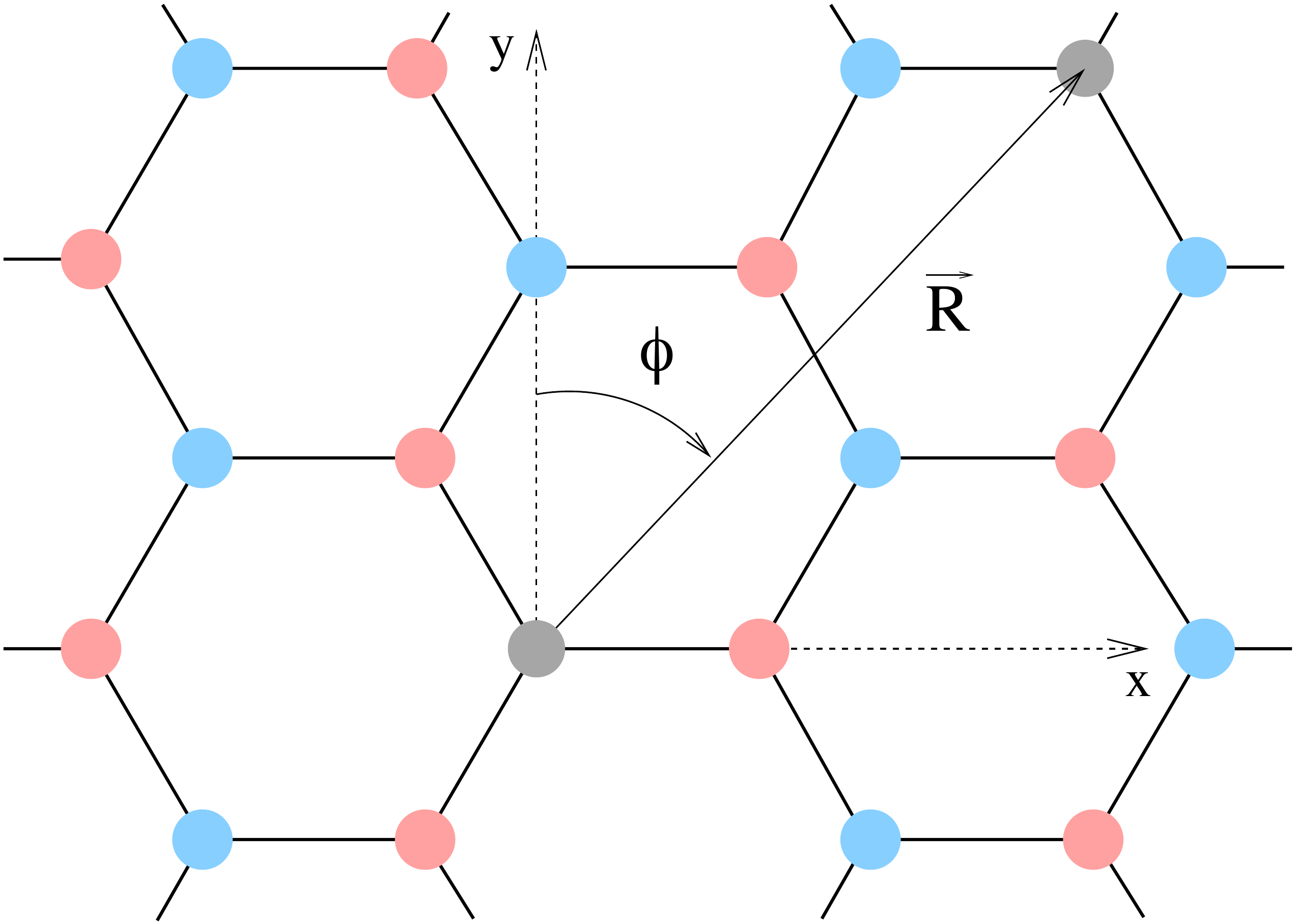}}
\caption{Graphene  inter-atomic distance is $a=0.14\, \rm nm$. Atoms  belonging to different sublattices (A and B) are  represented with different colors (red and blue).
 Two on-site impurities (grey circles) are placed on graphene ($AB$-configuration shown).  Periodic boundary
 conditions are assumed in the armchair ($x$) and zigzag ($y$) directions. The angle $\phi$ is counted from a zigzag direction
 (the $y$-axis).  }
\label{fig1}
\end{figure}
the dopants reside on the same ($AA$-case) or
opposite sublattices ($AB$-case). In the {\it same-sublattice} case the interaction is
repulsive,
   \begin{equation}
 \label{AA}
 W_{\! AA}({\bf R}) =\frac{\hbar \pi v \cos^2\!{\theta_{\! AA}}}{2R\ln^2\!{(R/a)}}.
 \end{equation}
The phase angle $ \theta_{\! AA}  ({\bf R}) = \frac{2\pi R}{3\!
\sqrt{3}a} \cos\phi$ depends on both the length of the inter-impurity
radius-vector ${\bf R}$ and the angle $\phi$ it makes   with a
zigzag direction, see Fig.~1.

The interaction between two impurities  residing on {\it
different}
sublattices is more interesting since it is  sensitive to the chemical potential $\mu$:
\begin{equation}
\label{AB}
W_{\! AB}({\bf R}) =\frac{\hbar \pi v \sin^2\!{\theta_{\! AB}}}{2R\ln^2\!{(R/a)}}-
 2\mathcal{E}_{\bf R} \Theta (\mathcal{E}_{\bf R}-|\mu|),
\end{equation}
where ${\theta_{\! AB}}({\bf R})=\frac{2\pi R}{3\! \sqrt{3}a}
\cos\phi+\phi$. The first  term in Eq.~(\ref{AB}) is repulsive.
Its origin is the same as that of $W_{AA}$: the band spectrum
of the conduction electrons is modified by the interaction with the
two impurities to an extent that depends on their relative
positions. The second term is the result of the formation of
two {\it bound states}, one above and one  below the Dirac point, at
$E =\pm \mathcal{E}_{\bf R}$, with the absolute value of the
energy given by\cite{LTM}
 \begin{equation}
 \label{AB level}
\mathcal{E}_{\bf R} = \frac{\hbar v |\sin{\theta_{\! AB}}|} {R\ln{(R/a)}}.
 \end{equation}

The second term in Eq.~(\ref{AB}) simply represents the energy
of an occupied bound state, $-2
\mathcal{E}_{\bf R}$, with the factor $2$ accounting for the spin degeneracy.
This term is logarithmically dominant (for $\ln{(R/a)}\gg 1$) over the first
term in Eq.~(\ref{AB}), but only when the chemical potential is confined between the upper and lower
bound states, $-\mathcal{E}_{\bf R}<\mu < \mathcal{E}_{\bf R}$.
 However, when the
chemical potential moves to either below or above the two
levels, both of them become empty or filled, respectively, and
their contribution to $W_{AB}$ disappears. This is accounted
for by the $\Theta$-function in Eq.~(\ref{AB}). The sign of the interaction of the two dopants
can thus be changed by a mere variation of the chemical potential\cite{LTM}.

This pairwise interaction of hydrogen atoms,
Eqs.~(\ref{AA})-(\ref{AB}), may have a profound effect on
the migration tendencies in favor of either clustering  or spreading, depending
on the chemical potential, which controls the attractive or repulsive nature of the
 interaction. It is thus important to evaluate the migration rate of hydrogen adatoms.

  Let us start with the propagation of a single
hydrogen atom on an ideal graphene crystal. Hydrogen atoms
reside above carbon atoms. Because of the large mass of
 the hydrogen atom (compared with the electron mass for example) the tunneling
 rate between carbon sites is rather small but not insignificant, and we are going to show that it would still result in a rapid spreading of the adatoms.

However, we are also going to see that the presence of a second adatom, even many interatomic distances
away, changes the situation dramatically. Indeed, the difference of the
energies before and after tunneling, $\Delta E$, is of the order of
several to tens of $\rm meV$. This energy is many orders of magnitude
larger than the broadening of the hydrogen level due to elastic tunneling.
Therefore, in order to occur, a tunneling event must
be assisted by some mechanism susceptible to deliver or carry away the energy difference.  The two candidates for this are phonons and electrons, so we are going to calculate the rate of
such phonon-assisted  and electron-assisted tunneling of hydrogen on graphene.

Throughout the paper we utilize the effective Dirac model of
graphene spectrum applicable at low energies. Correspondingly,
our results are valid when typical distances between adatoms
are much larger than the carbon lattice spacing. For such long
distances the numerical methods, such as DFT, would be
impractical. On the other hand, our approach cannot be applied
for short distances, in particular to the problem of hydrogen
dimers where numerical methods are needed\cite{Casolo,Slji}.

The plan of the paper is as follows. In Section \ref{hhopping}
we review the hopping of a single  hydrogen adatom on graphene.
We find the inelastic tunneling rates for the processes
involving phonons and electrons, in Sections \ref{phonon} and
\ref{electron}  respectively. Section \ref{discuss} is a
discussion of the results.

\section{Hydrogen hopping on graphene\label{hhopping}}
First-principle calculations of the height of the potential
barrier   between first neighbors for hydrogen adatoms resulted
in a broad range of
values\cite{ferro,ferrobis,hornekaer,chen,lei,wehling,boukhvalov,mckay,huang,BORO}
from $0.29\,{\rm eV} to 1.3\,{\rm eV}$. This is a consequence of
the relatively small size of the system simulated by DFT
combined with the important role played by the carbon lattice
relaxation in the binding of adatoms.

Coronene and coronene-like species have been shown to
effectively simulate the electronic structure of graphene
hydrogenation, especially near central carbon sites
\cite{Irle}. Using a development version of the Q-Chem 4.3
quantum chemistry package \cite{Qchem}, we modeled the
potential landscape of a hydrogen adatom above a central carbon
atom of circumcoronene ($C_{54}H_{18} + H$). All DFT
calculations were carried out on the doublet electronic ground
state using the B3LYP functional with the double-zeta polarized
basis set 6-31G(d).

\begin{figure}[h]
\resizebox{.50\textwidth}{!}{\includegraphics{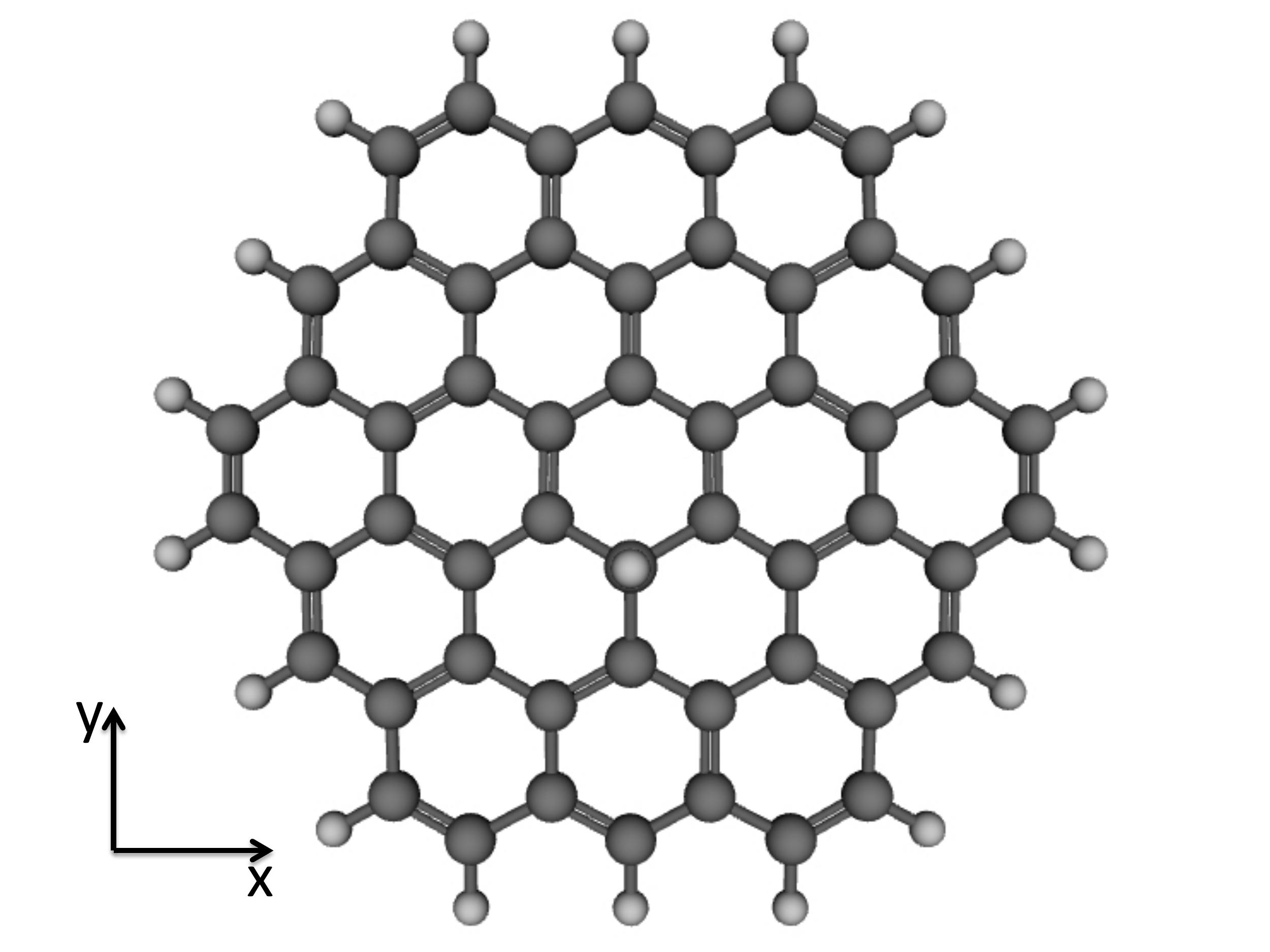}}
\caption{Hydrogenated circumcoronene ($C_{54}H_{18} + H$), the adsorption site is shown for reference.}
\label{fig2}
\end{figure}

A full geometry optimization with the adatom was first computed
allowing carbon  lattice relaxation. After lattice relaxation,
DFT results show an in-plane carbon-carbon distance
$a=0.14\,{\rm nm}$ with a torsion angle of $\theta = 15$
degrees. The hydrogen adatom was then scanned along an
evenly-spaced square $0.7\,{\rm nm}$ x $0.7\,{\rm nm}$ grid
above the carbon atom allowing relaxation of the z-component of
the gradient for the hydrogen (see Fig.~\ref{fig2}) and
constraining the carbon lattice to that of the optimized
structure. The minimum potential energy was taken at each step.

\begin{figure*}[htp]
  \centering
  \subfigure[ Potential energy surface $U_{DFT}^{'} (x,y)$]{\includegraphics[scale=0.65]{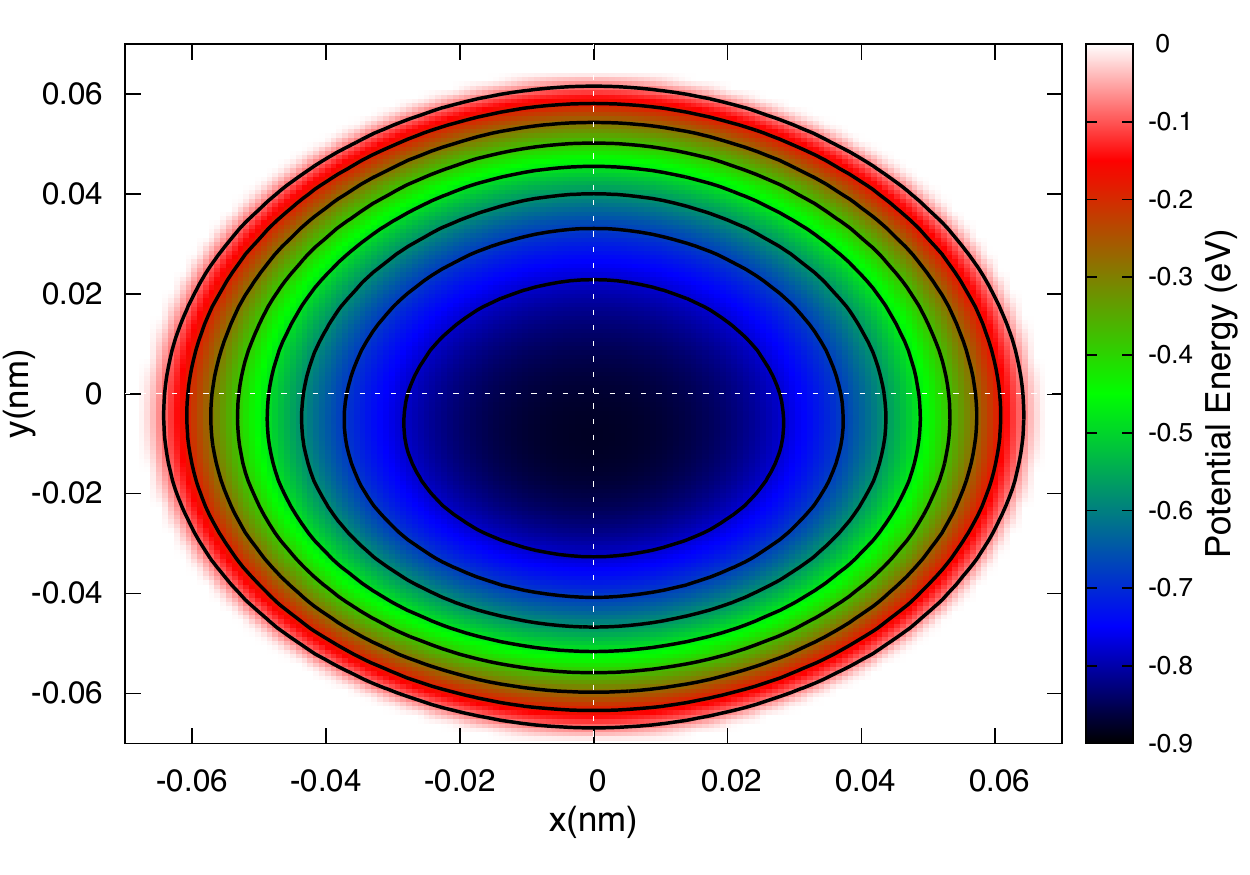}}\hfill
  \subfigure[ Ground state wave function]{\includegraphics[scale=0.65]{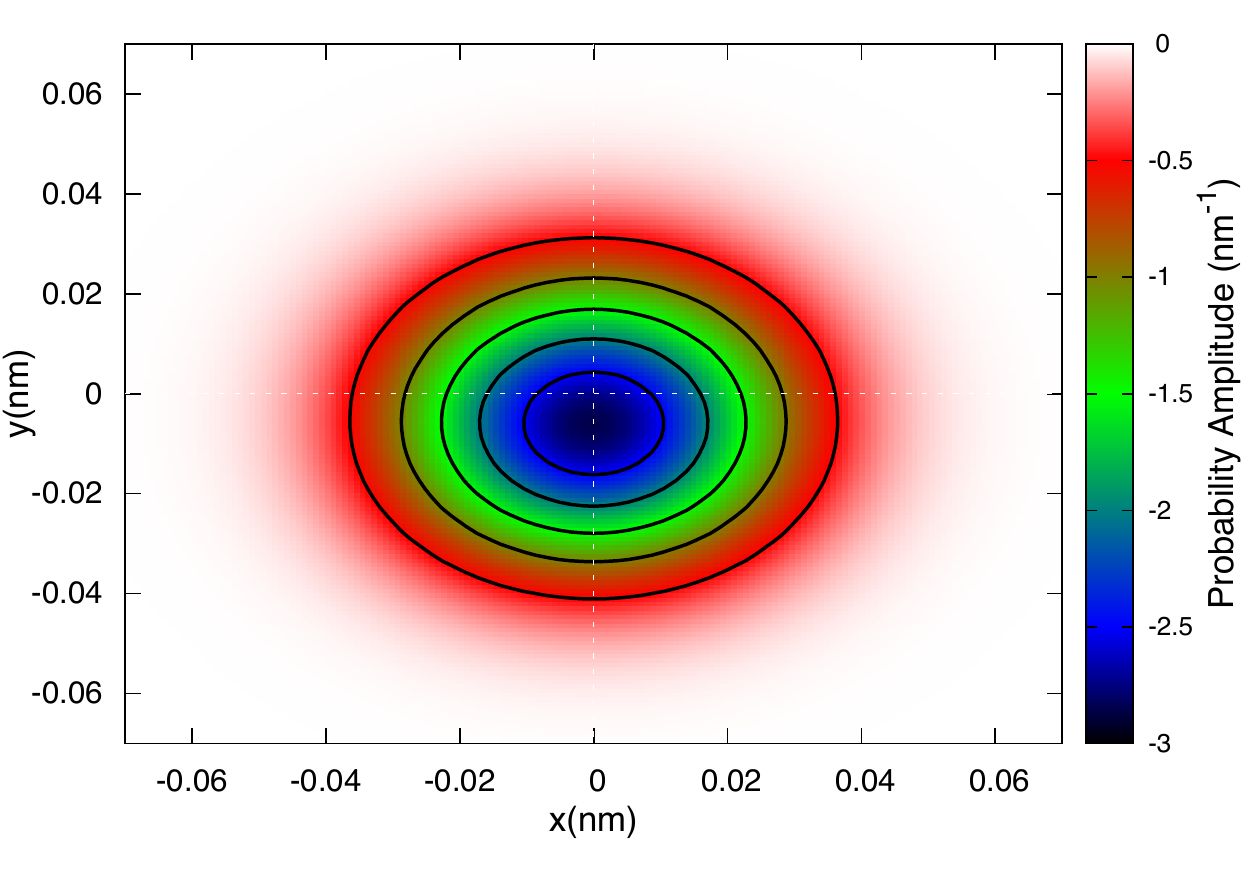}}
  \caption{$U_{DFT}^{'} (x,y)$ (a) and the ground state wave function (b) of a single hydrogen adatom above one of the central carbon sites of circumcoronene. The potential energy $U_{DFT}^{'} (x,y)$ was restricted to a disk of radius $a/2$ outside of which the energy was kept constant.  The Shr\"odinger equation was then solved in a domain extending to  the regions of the nearest neighbors. Only the region of the central carbon site is shown here. The contour lines in (a) are equally spaced by $0.1\,{\rm eV}$ while in (b) they are equally spaced by  $0.5\,{\rm nm^{-1}}$.}
  \label{fig3}
\end{figure*}

In order to get an approximation for the hopping amplitude, we
truncated  the potential  obtained from DFT to within an in-plane disk
of radius $a/2$, keeping the potential constant outside of this region. 
This allows us to set up the following Hamiltonian in the
position representation
\begin{equation}
 \label{Ham}
\hat{H} = -\frac{\hbar^2}{2 M_{H}} \left(\frac{\partial^2 }{\partial x^2} + \frac{\partial^2 }{\partial y^2} \large \right) + U_{DFT}^{'} (x,y)
\end{equation}
where $M_H$ is the mass of a hydrogen atom and $U_{DFT}^{'}
(x,y)$ is the  truncated DFT potential.
The ground state wave function of the 
Hamiltonian~(\ref{Ham}) extended to large distances $r\gg a/2$ was obtained using the Fourier grid
Hamiltonian method \cite{Marston}. This gives us an approximation to the evanescent tail of the wave function in that region. Due to the large size of the
grid, diagonalization was done numerically using the Primme
iterative diagonalization routine \cite{primme}. The results
are shown in Fig.~\ref{fig3}.

Using this model of a ground state wave function, we
numerically estimate the  hopping amplitude $t = -0.61\,\mu
{\rm eV}$ by integrating the product of two neighboring well
ground state wave functions multiplied by the DFT potential
over a circular region of radius $a/2$ centered on one of the
two sites. At the same time, we can estimate the overlap
integral for the wave functions of hydrogen states residing on
adjacent atoms:  $I_{{\bf R}-{\bf R'}}=\int d{\bf r}\psi^*({\bf
r}-{\bf R})\psi({\bf r}-{\bf R'})$, which plays a determinant
role in the evaluation of the transition rate. Using the same
model for the ground state wave function as above, numerical
integration gives $I_{{\bf R}-{\bf R'}} = 3.8 \times 10^{-6}$.
This approximation of $I_{{\bf R}-{\bf R'}}$ is an
underestimate by construction of the model ground state wave
function, as it ignores the other wells in the lattice.
However, the small values of both $t$ and $I_{{\bf R}-{\bf
R'}}$ are reflecting the sharply localized nature of the
hydrogen wave functions illustrated in Fig.~\ref{fig3}, a
posteriori justifying the approximation made.

A {\it single} hydrogen atom hopping over an ideal graphene sheet must therefore behave like a band
particle and propagate with the band velocity $u\sim at/\hbar\sim 1~\text{cm/s}$. This velocity is much smaller than the band velocity of
$\pi$-electrons, $v= 1 \times 10^6$ m/s. Nonetheless, if hydrogen atoms absorbed by graphene were moving with such velocity $u$, this would result in a rather quick homogenization of
their distribution.

However, such a small (by microscopic scales) velocity means that the hydrogen bandwidth is very narrow. This should cause the suppression of
elastic hopping as soon as the on-site potential energies between neighboring sites are different by more
than $t$.  This is why the interaction mediated by conduction
electrons makes elastic hopping {\it impossible}.  When one of the atoms hops, it virtually always {\it changes} the sublattice, $A \to
B$ or $B\to A$: the likelihood of tunneling to a second-nearest neighbor on the same sublattice site is negligible, because of
the much larger distance that must be covered.  Suppose now that a second hydrogen atom
happens to be some distance ${\bf R}$ away, sitting atop a carbon atom belonging to the same sublattice. The
interaction between the atoms then results in a {\it positive} energy $W_{AA}({\bf R})>0$. When one of the
atoms hops to its nearest neighbor carbon atom, it lands on the opposite sublattice, and the interaction
energy in the final state is {\it negative}, $W_{AB}({\bf R}+{\bf a})<0$. The total energy change in this
process, $W_{AA}({\bf R})-W_{AB}({\bf R}+{\bf a})$, therefore, exceeds both $W_{AA}$ and $W_{AB}$.
Even for two atoms sitting as far as a thousand interatomic distances apart, the energy change is estimated  from
Eqs.~(\ref{AA})-(\ref{AB}) to be of the order of $1 \,\rm meV$. This energy is many orders of
magnitude greater than the hydrogen adatom hopping bandwidth. We conclude, therefore, that the interaction mediated by conduction
electrons makes elastic hopping impossible and results in a collective pinning of hydrogen on graphene.

Consequently,  the tunneling can  occur only if facilitated by  additional processes which supply
(or carry away) the  energy difference. Such processes must involve excitations of graphene,
phonons or electron-hole pairs. The transition rates in both cases are evaluated in Sections \ref{phonon} and \ref{electron}, respectively.

\section{Hydrogen-phonon interaction\label{phonon}}

Let us consider inelastic tunneling of a hydrogen atom between adjacent sites ${\bf R}$ and ${\bf R}'$ where the atom has potential energy $W_{\bf R}$ and
$W_{{\bf R}'}$, respectively. The extra energy $W_{\bf R}-W_{{\bf R}'}$, depending on its sign, must be carried away or supplied  by phonons.

There are three low-energy phonon modes in graphene:
longitudinal, transverse, and flexural. In the long wavelength limit, the first two have
the usual acoustic dispersion, $\omega_q =s_{l,t}q$, while the third one has a
quadratic dispersion. For the same frequency, long-wavelength flexural
phonon modes have a much larger wave-vector and, accordingly, reside in a
much larger phase space. We expect flexural modes to dominate
phonon-assisted processes, similarly to how they dominate the phonon resistivity in
electron transport\cite{MO}.

The flexural phonon's quadratic spectrum follows from the Hamiltonian for the out-of-plane sheet oscillations, which, in the long-wavelength limit, is\cite{CGP}
\bq
\label{flexural Hamiltonian}
H_0=\frac{1}{2}\int d^2 r \Bigr[ \rho\dot h^2+{D}(\nabla^2 h)^2\Bigr],
\ee
with $\rho$ and $D$  denoting the two-dimensional mass density and the flexural rigidity constant of graphene, respectively. The Hamiltonian (\ref{flexural Hamiltonian})  describes unstrained graphene. In the presence of strain the Laplacian in $H_0$ is replaced with $\nabla h$, in which case $H_0$ assumes the form of the usual membrane Hamiltonian. Below, we consider an unstrained sheet. As follows from the Hamiltonian (\ref{flexural Hamiltonian}), the spectrum of flexural modes is
 \bq
 \label{flexural spectrum}
 \omega_q=\sqrt{\frac{D}{\rho}}\,q^2.
 \ee

Standard quantization of the Hamiltonian (\ref{flexural Hamiltonian}) leads to the following expression for the phonon flexural disturbance
\begin{eqnarray}
\label{phonons quantized}
  h({\bf r})=\sum_q \sqrt{\frac{\hbar}{2\mathcal A\rho\omega_q}}\left(  a_{\bf q} e^{-i\omega_q t+i{\bf q}\cdot{\bf r}}+h.c.\right),
\end{eqnarray}
in terms of the phonon creation and annihilation operators; $\mathcal A$ stands for the normalization area of the sheet.

The hydrogen adatom represents a mass defect stuck to the graphene lattice. The coupling of the atom to the phonon field (\ref{phonons quantized}) is described by the additional kinetic energy of the atom's oscillatory motion induced by the phonons,
\bq
\label{H-phonon coupling}
\mathcal H_{H-ph}=\frac{M}{2} {\dot {  h}}^2({\bf R}).
 \ee
 Substitution of the phonon operator (\ref{phonons quantized}) into this expression yields the hydrogen-phonon coupling Hamiltonian
\bq \label{H-phonon quantized} \mathcal H_{H-ph}=\mathcal
H_{abs}+\mathcal H_{em}+\mathcal H_{sc},
\ee
which consists of
three terms. The first term describes absorption of two phonons
with wave-vectors ${\bf q}$ and ${\bf k}$,
\bq
\label{absorption
hamiltonian} \mathcal H_{abs}=-\frac{M}{2} \sum_{{\bf q},{\bf
k}}A_qA_{k}  a_{\bf q}   a_{\bf  k}
e^{-i(\omega_q+\omega_{k})t+i({\bf q}+{\bf k})\cdot {\bf R}},
\ee
where $A_q=\sqrt{\hbar \omega_q/2\rho \mathcal A}$. The
second term is the Hermitian conjugate to the first one,
\bq
\label{emission hamiltonian} \mathcal H_{em}= -\frac{M}{2}
\sum_{{\bf q},{\bf k}}A_qA_{k}   a_{\bf q}^\dag   a_{\bf
k}^\dag e^{i(\omega_q+\omega_{k})t-i({\bf q}+{\bf k})\cdot {\bf
R}},
\ee
and describes the emission of two phonons. Finally, the
term \bq \label{scattering hamiltonian} \mathcal H_{sc} = M
\sum_{{\bf q},{\bf k}}A_qA_{k}  a_{\bf q}   a_{\bf
k}^\dag e^{-i(\omega_q-\omega_{k})t+i({\bf q}-{\bf k})\cdot{\bf
R}}, \ee describes scattering processes, in which a phonon with
wave-vector ${\bf q}$ is absorbed and a phonon with wave-vector ${\bf k}$ is
emitted.

In what follows we are going to utilize the perturbation theory
to take into account the hydrogen-phonon coupling
(\ref{H-phonon quantized}). Such approach is justified because
of the light mass of hydrogen $M \ll M_C$, which ensures that
the coupling is going to be rather weak.

The probability of phonon-assisted hydrogen hopping from site
${\bf R}$ to site ${\bf R}'$ follows from the Golden rule. For
example, when hopping occurs with a decrease of the on-site
energy, $\hbar \omega_0 =W_{\bf R}-W_{\bf R'}>0$, only phonon
emission and scattering processes are allowed. The transition
rate for the emission is
\begin{equation}
 \label{emission rate}
W_{em}=\frac{2\pi}{\hbar}|\langle {\bf R}';{\bf q},{\bf k}|\mathcal
H_{em}|{\bf R}\rangle|^2\delta(\hbar \omega_0 -\hbar \omega_q -\hbar \omega_k).
\end{equation}
From the Hamiltonian (\ref{emission hamiltonian}), we obtain
that the matrix element for the emission transition is
\bq
\label{emission matrix element}
|\langle {\bf R}';{\bf q},{\bf k}|\mathcal
H_{em}|{\bf R}\rangle|^2=\frac{\hbar^2 M^2I^2_{{\bf
R}-{\bf R'}}}{16 {\mathcal A}^2\rho^2} \,\omega_q \omega_{k}(1+N_q)(1+N_k)
\ee
where $N_q = (e^{\beta \omega_q}-1)^{-1}$ is the
Bose-Einstein distribution with  $\beta=\hbar/k_BT$.

The total  emission rate is found from Eqs.~(\ref{emission rate})-(\ref{emission matrix element}) by integrating over all
phonon momenta. After simple calculation one finds,
\begin{equation}
\label{total probability of absorption}
W_{em}  =\frac{M^2I_{{\bf R}-{\bf R'}}^2}{128\pi\rho D}\int\limits_0^{{\omega_0}} d\omega\,
\frac{\omega}{1-e^{-\beta \omega}}\,\frac{{\omega_0}-\omega}{1-e^{-\beta({\omega_0}-\omega)}}.
\end{equation}

Similarly, the probability of hydrogen atom hopping assisted by the absorption of a phonon with frequency $\omega$  and emission of a phonon with the higher frequency $\omega_0+\omega$ is given by
\begin{equation}
\label{total probability of scattering}
W_{sc}  =\frac{M^2I_{{\bf R}-{\bf R'}}^2}{32\pi\rho D}\int\limits_0^{\infty} d\omega\,
\frac{\omega}{e^{\beta \omega}-1}\,\frac{{\omega_0}+\omega}{1-e^{-\beta({\omega_0}+\omega)}}.
\end{equation}

The transition rates (\ref{total probability of absorption}) and (\ref{total probability of scattering}) can be easily calculated in the limits of low and high temperatures. It turns out that the scattering channel is the dominant mechanism of phonon-assisted hopping when temperatures are high. On the other hand, emission prevails at low temperatures. This is expected since the number of phonons available for scattering is small in that case.

\subsection{High temperatures, $\beta\omega_0\ll 1$}

In this limit,  the denominators in the emission probability (\ref{total probability of absorption}) can be expanded to
the linear order in small $\beta$. In the scattering probability (\ref{total probability of scattering}) one may set $\omega_0=0$ since
typical phonons participating in the process have much higher frequency, $\omega \sim 1/\beta \gg \omega_0$.
  Using the identity, $\int_0^{\infty}du\,{u^2}/{\sinh^2(u/2)}={4\pi^2}/{3}$, we obtain,
\begin{equation}
\label{high temperatures}
\left. \begin{array}{l} W_{em}\\ W_{sc} \end{array} \right\} = \frac{M^2I_{{\bf R}-{\bf R'}}^2 }{32\pi \rho D} \left(\frac{k_B T}{\hbar}\right)^2
\left\{ \begin{array}{l}  \omega_0/4,\\ \pi^2 k_B T/3\hbar.
\end{array} \right.
\end{equation}

As mentioned above, at high temperatures the scattering-assisted processes are much more efficient in facilitating tunneling than the emission transitions.
The $T^3$-dependence of the scattering-assisted probability can be understood as follows. Phonons involved in the transition have energy of the order  $k_BT$. One power of temperature arises from the number of such phonons available. The other two powers appear due to the fact that high-frequency phonons interact  stronger with a hydrogen adatom, since they result in a larger coupling Hamiltonian (\ref{H-phonon coupling}).

\subsection{Low temperatures, $\beta\omega_0\gg 1$}

In this limit,  all  exponentials with negative arguments can be discarded in both Eqs.~(\ref{total probability of absorption}) and (\ref{total probability of scattering}).
 Additionally, since in the scattering channel the incident phonons have frequency $\omega \sim 1/\beta \ll \omega_0$, one
 may neglect $\omega$ in the numerator of the integrand of Eq.~(\ref{total probability of scattering}). Using the fact that
 $\int_0^{\infty}du\,{u^2}/(e^u-1)={\pi^2}/{6}$, we find,
\begin{equation}
\label{low temperatures}
\left. \begin{array}{l} W_{em}\\ W_{sc} \end{array} \right\} = \frac{ M^2I_{{\bf R}-{\bf R'}}^2 }{192\pi\rho D}\, \omega_0
\left\{ \begin{array}{l}  \omega_0^2/4,\\ (\pi k_B T/\hbar)^2.
\end{array} \right.
\end{equation}
 The predominance of emission over scattering in the low-temperature limit is driven by the scarcity of phonons in the initial state.

 \subsection{Hopping with the increase in energy, $\omega_0<0$}

 When a hydrogen atom tunnels from a site with a lower interaction energy $W_{\bf R}$ to a site with a higher energy $W_{{\bf R}'}$ it needs to pick up the extra energy to do so. This is possible by either absorbtion of two phonons via the process described by the Hamiltonian (\ref{absorption hamiltonian}) or via the  scattering processes already discussed above. The corresponding rates can be calculated from the same Golden rule formalism. However, they also follow immediately from the detailed balance principle, which provides,
 \begin{eqnarray}
 \label{detailed balance}
 W_{abs}(-\omega_0)=W_{em}(\omega_0) e^{-\beta\omega_0},\nonumber \\ W_{sc}(-\omega_0)=W_{sc}(\omega_0) e^{-\beta\omega_0}.
 \end{eqnarray}

At high temperatures, $\beta \omega_0 \ll 1$, therefore, the transition rates for the upward (in energy) tunneling are virtually the same as for the downward transitions. At low temperatures, $\beta \omega_0 \gg 1$, however, the upward transitions are strongly suppressed.

\section{Hydrogen-electron interaction\label{electron}}

Inelastic tunneling of a hydrogen atom can also occur with the
excess (or deficit of) energy transferred to (from) the
conduction electrons. At low energies, the latter have the Dirac spectrum   $\epsilon=\pm
vp$.

The amplitude of this
process is proportional to the matrix element of the
interaction of a hydrogen atom with the electric charge of a
conduction electron. A hydrogen adatom may interact with conduction electrons via
Coulomb forces. Since the adatom is neutral, this
interaction, in the lowest order, results from the dipole moment associated with the
carbon-hydrogen bond and has the form,
\begin{equation}
\label{dipole-charge hamiltonian}
  H' = - \frac{e \, {\bf d}_\parallel\cdot ({\bf r}-{\bf R})}{\kappa |{\bf r}-{\bf R}|^3},
\end{equation}
where ${\bf r}$ is the coordinate of the electron and $ {\bf
d}_\parallel$ is the adatom's in-plane dipole moment. The
coefficient $\kappa$ is the effective dielectric constant of
graphene, which describes screening of static electric fields
by other conduction electrons.

From symmetry considerations, however, it
is clear that the {\it average} dipole moment will be
pointed in the direction {\it perpendicular} to the plane of
graphene and thus the expectation value of the Hamiltonian (\ref{dipole-charge
hamiltonian})  in the ground state of a hydrogen adatom is zero,
$\langle 0|   H'|0 \rangle =0$.  As long as there is no  spontaneous
symmetry breaking, we need to explore the
second order correction.

This second order correction is analogous to the van der Waals
interaction between two atoms where the interaction arises as a
result of quantum fluctuations. The main difference comes from
the fact that in our situation one of the particles (the electron) has a net charge,
so that the resulting interaction falls of as $1/r^4$, rather
than as $1/r^6$. Indeed,  the second order energy correction is finite and given by the standard
expression,
\begin{equation}
\label{second order}
\Delta E = \sum_{n} \frac{|\langle n|   H'|0 \rangle |^2}{E_0 - E_n},
\end{equation}
where the summation is taken over all excited states $n$.

The actual energy levels and dipole moment's matrix elements
for a hydrogen atom sitting above a carbon atom in a graphene
crystal can only be determined by first-principles
calculations which are beyond the scope of the present paper.
However, it is not difficult to estimate the correction
(\ref{second order}) by the order of magnitude. The
non-diagonal matrix elements are $\langle n| {\bf d}_\parallel
|0 \rangle \sim e a_B$, where $a_B$ is the Bohr radius.
Similarly, the difference $E_0 - E_n$ is of the order of the
Rydberg energy, $e^2/2a_B$. As a result, $\Delta E$, which
should be identified with the effective hydrogen-electron
coupling energy, is,
\begin{equation}
\label{effective energy}
V({\bf r}-{\bf R}) \equiv \Delta E = -{\cal C} \frac{e^2 a_B^3}{\kappa |{\bf r}-{\bf R}|^4}.
\end{equation}
where ${\cal C}$ is an unknown positive dimensionless constant.

Expression (\ref{effective energy}) can also be obtained
by purely classical considerations. The electric field produced by
the conduction electron, and acting on the hydrogen atom, is
$e/\kappa r^2$; for simplicity we set ${\bf R}=0$. This field is
$(r/a_B)^2$ times weaker than the typical atomic fields,
$e/\kappa a_B^2$. It, therefore, leads to the displacement of the
atomic electron of the order, $a_B (a_B/r)^2$, resulting in the
induced dipole moment $d_\parallel \sim e a_B^3/r^2$, from
which Eq.~(\ref{effective energy}) follows. Note that, for a
free hydrogen atom, the dimensionless constant is
known exactly\cite{LLIII}, ${\cal C} =9/2$, suggesting that ${\cal C} \sim
1-10$ when the atom is sitting atop a graphene sheet.

We can now write the hydrogen-electron interaction
(\ref{effective energy}) in the second-quantized form with the
help of the electron creation and annihilation operators,
\begin{eqnarray}
\label{h-e second quantized}
  H_{H-e} = \frac{1}{2{\cal A}} \sum_{{\bf p},{\bf p}'} \sum_{\alpha, \beta} V({\bf p}-{\bf p}') e^{i({\bf p} -{\bf p}')
\cdot {\bf R}/\hbar}\nonumber \\ \times \,   c^\dagger_{\alpha \bf p}   c_{\beta {\bf p}'} [1+\alpha \beta\cos{({\bf p},
{\bf p}')}].
\end{eqnarray}

In this expression the indices $\alpha$ and $\beta$
assume two values: $+1$ for electrons in the  upper Dirac cone
and $-1$ in the lower cone; the trigonometric factor in the
brackets of the second line is coming from the pseudospin
projection of the initial electron state $|\beta {\bf p}'
\rangle$ onto the final state  $|\alpha {\bf p}\rangle$.

The Fourier transform of the interaction potential diverges at low distances and should be cut-off there by the Bohr radius,
\begin{eqnarray}
\label{interaction Fourier}
V(\hbar {\bf q}) = \int d^2r e^{-i{\bf q}\cdot {\bf r}}V({\bf r}) \approx - 2\pi {\cal C}\frac{e^2 a_B^3}{\kappa}
\int\limits_{a_B}^\infty \frac{dr}{r^3} J_0(qr) \nonumber\\ =-\pi {\cal C} \frac{e^2 a_B}{\kappa}.~~~~
\end{eqnarray}
The interaction is essentially short-range with the strength
independent of the electron momenta. This is the result of the
fast decay of the interaction (\ref{effective energy}) with the
distance and the fact that only long-wavelength electrons,
$p\,a_B \ll \hbar$, can participate in the interactions, since
energy involved in inelastic tunneling processes, $\hbar
\omega_0$, is small compared with the electron bandwidth.

The rate of inelastic tunneling transitions facilitated
by the electron scattering is given by the Golden rule formula
with the matrix element provided by Eq.~(\ref{h-e second
quantized}):
\begin{eqnarray}
\label{probability electron general}
W_{H-e}=\frac{3\pi {\cal C}^2 e^4 a_B^2}{4\hbar^5\kappa^2} I_{{\bf R}-{\bf R'}}^2 \sum_{\alpha\beta} \int\limits_0^\infty pdp \int\limits_0^\infty kdk  \, (1-n_{\alpha p}) n_{\beta k} \nonumber \\ \times \delta (\hbar \omega_0 -\alpha vp +\beta vk),~~~~
\end{eqnarray}
where $n_{\alpha p}=[\exp{(\alpha vp/T)+1]^{-1}}$ is the
Fermi-Dirac distribution of electrons in the two cones. The overall coefficient
in Eq.~(\ref{probability electron general}) takes into account
the existence of two Dirac cones as well as the two-fold spin
degeneracy. Note that $[1+\alpha \beta\cos{({\bf p},{\bf
p}')}]^2$ yields the factor $3/2$ upon the angle averaging.

At high temperatures, $k_BT \gg \hbar \omega_0$, the main
contribution comes from transitions that occur within the same
cone, $\alpha =\beta$. In this limit $\hbar \omega_0$ can be
neglected. After a simple integration we obtain,
\begin{equation}
\label{probability electron high T}
W_{H-e}=\frac{\pi^3 {\cal C}^2 e^4 a_B^2 k_B^3 T^3}{4\hbar^5\kappa^2 v^4} I_{{\bf R}-{\bf R'}}^2.
\end{equation}

At low temperatures, $k_BT \ll \hbar \omega_0$, the dominant
process results from the lifting of an electron from the lower
cone, $\beta =-1$, into the upper cone, $\alpha =1$, yielding
the following inelastic tunneling rate,
\begin{equation}
\label{probability electron low T}
W_{H-e}=\frac{\pi {\cal C}^2 e^4 a_B^2 \omega_0^3}{8\hbar^2\kappa^2 v^4} I_{{\bf R}-{\bf R'}}^2.
\end{equation}

Comparing these results with the rates for phonon-assisted
tunneling from Section \ref{phonon}, we see that the electron
mechanism leads to the same temperature dependence as the
dominant phonon processes: phonon scattering at high
temperatures, see Eq.~(\ref{high temperatures}), and phonon
emission at low temperatures, see Eq.~(\ref{low temperatures}).
The reason for this coincidence is not difficult to understand.
For example, at high temperatures electrons participating in
the tunneling are confined in a smaller region of phase space; however, their
coupling to hydrogen is stronger, since the coupling
Hamiltonian  (\ref{h-e second quantized}) remains constant in
the long-wavelength limit while coupling to phonons,
Eq.~(\ref{scattering hamiltonian}), vanishes there.

\section{Discussion\label{discuss}}

Because of the similar temperature dependence, the  ratio of
the two rates takes a rather simple form. Interestingly, the
two mechanisms appear to lead to rates of the same order of magnitude. At high temperatures, $k_BT \gg \hbar \omega_0$, the
ratio of the two rates can be written as
\begin{equation}
\label{estimate}
\frac{W_{H-e}}{W_{em}} = 24\pi^2 {\cal C}^2 \left(\frac{e^2}{\hbar\kappa v}\right)^2 \left( \frac{\rho a_B^2}{M}\right) \left(\frac{D}{Mv^2} \right).
\end{equation}

The first ratio in this expression is $e^2/\hbar\kappa v =1.4$,
given the well-known value of the effective dielectric constant
in graphene\cite{CGP}, $\kappa = 1+e^2/4\hbar v=1.55$, and the Fermi
velocity, $v=1\times 10^6$ m/s. The second ratio is of the same
order, ${\rho a_B^2}/{M}= 16a_B^2/(\sqrt{3} a^2)=1.3$, and is easily calculated from
the value of the lattice spacing, $a=0.14$ nm,
 and the fact that the graphene lattice
is honeycomb. Given the generally quoted value for the flexural rigidity constant\cite{LIND} $D=1.2\,\rm eV$, the last ratio is very small, ${D}/{Mv^2}
=1.4\times 10^{-4}$. However, this smallness is compensated by the large value of the numerical prefactor in
Eq.~(\ref{estimate}). The constant ${\cal C}$ is not
known. However, if we utilize its  value for a free hydrogen
atom in the ground state\cite{LLIII}, ${\cal C}=4.5$, we obtain
that the ratio of the electron to the phonon contributions is
 \begin{equation}
\label{estimate 1}
\frac{W_{H-e}}{W_{em}} \approx 1.45
\end{equation}

At low temperatures, $k_BT \ll \hbar \omega_0$, the  ratio is
given by the  expression similar to Eq.~(\ref{estimate}) where
there is an additional coefficient $4$, as follows from
Eq.~(\ref{probability electron low T}) and the first line of
Eq.~(\ref{low temperatures}). As a result, we have
$W_{H-e}/W_{em} \approx 5.83$ there.

The rates for inelastic hydrogen tunneling obtained in the
preceding sections are rather small, predicting a very slow diffusion of hydrogen. Let us estimate the rate of hydrogen
hopping in the dominant regime.
Namely, consider the case of high temperatures where the
scattering channel dominates, and the transition rate is given
by the second line of  Eq.~(\ref{high temperatures}). Using
again\cite{LIND} $D=1.2\,{\rm eV}$, and the overlap integral calculated earlier, $I_{{\bf R}-{\bf R'}} \approx 3.5 \times 10^{-6}$, we
find that in the high temperature limit the total hopping rate provided by the electron-assisted and phonon scattering rates is  $W=W_{H-e}+W_{sc}  \approx 1.2 ({T}/{T_{0}})^3\rm\,Hz$ in terms of the reference temperature $T_{0}=300\,{\rm K}$.

It is also instructive to restate the hopping rate $W$
in terms of the diffusion coefficient ${\cal D}$.
To estimate its order of magnitude  we assume that $k_BT$ exceeds $\hbar |\omega_0|$ on each
site, so that the probability of tunneling $W$ is uniform across the extent of the entire graphene sheet and is set by the temperature $T$.
The diffusion
 coefficient for a particle hopping on a honeycomb lattice with
 a given probability $W$ is then,
 \bq
 \label{diffusion coefficient}
\mathcal D=\frac{1}{2}\,Wa^2.
 \ee
At room temperature the diffusion coefficient is of the order
${\cal D}\sim 10^{-21}$ m$^2$/s.  Given  this value of the
diffusion constant, it would take a hydrogen atom a time
$\tau \sim L^2/{\cal D} \sim 100$ years to diffuse across a
micron-sized graphene flake.

In this paper we disregarded polaronic effect. The latter could
be expected to renormalize mobility of a moving particle, by
some numerical factor.  On the other hand, the suppression of
diffusion due to the effective interaction studied here appears
to be a much stronger effect, by orders of magnitude.

Because of the slow character of hydrogen adatom diffusion, hydrogen desorption and absorption might become important in specific situations. Those processes depend on the ambient conditions and are beyond the scope of this paper.

\section{Summary}

For electronics applications utilizing graphene  samples doped with
resonant adatoms, such as hydrogen, it is important to predict
how fast the adatom diffusion would occur at room temperatures.
Fast diffusion could be favorable for the device manufacturing,
but it would be detrimental for the device longevity. On the
other hand, fast diffusion would be required for the
realization of various phases resulting from adatom
ordering\cite{CSA,ASL,KCA}.

Our calculations show that the diffusion of hydrogen adatoms, the lightest dopants, is very slow. A single adatom placed on an ideal graphene sheet would propagate with the velocity of the order of 1\,cm/s.  However, the presence of other such adatoms suppresses adatom propagation dramatically. This is the result of a peculiar electron-mediated effective interaction between adatoms. Two features make this interaction efficient in suppressing diffusion. First, it is a is  long-range interaction, so that it is considerable even when adatoms are tens of nanometers away from each other. Second, it changes sign depending on whether adatoms reside on the same or opposite sublattices. In order to move, an adatom would have to hop to a nearest neighbor carbon site and change its energy in the process  by as much as tens of meV. Such inelastic hopping requires assistance from phonons or electron-hole excitations.

Let us emphasize that the suppression of the diffusion
discussed in this paper is due to the graphene band structure
featuring two Dirac points, which  results in a quantum
interference of the electron band states propagating over two
sublattices in the honeycomb arrangement of carbon atoms. The
effective hydrogen-hydrogen interaction, which has opposite
sign on the two sublattices, originates from coupling of
hydrogen adatoms to such band electrons.

As a result, we find a single adatom hopping time to be in the
range of milliseconds at room temperature. This indicates that
the diffusion is sufficiently slow that hydrogen-doped graphene
devices are feasible. However, it is too slow for any
gate-control of the hydrogen adatom distribution to be useable
in  fast switching devices.

\acknowledgments
We thank Oleg Starykh and Janvida Rou for helpful discussions.

S.L. and J.T. acknowledge the support by NSF through MRSEC DMR-1121252.
E.M. was supported by the Department of Energy, Office of Basic
Energy Sciences, Grant No.~DE-FG02-06ER46313.

\end{document}